\documentclass[useAMS,usenatbib]{mn2e}
\usepackage{float,multicol,epsfig}
\usepackage{amssymb}

\title[Comment on `Recombination induced softening and reheating
of the cosmic plasma']
{Comment on `Recombination induced softening and reheating
of the cosmic plasma'}

\author[W. Y. Wong and D. Scott]{Wan Yan Wong\thanks{E-mail:
wanyan@phas.ubc.ca}
and Douglas Scott\thanks{E-mail: dscott@phas.ubc.ca} \\
Department of Physics and Astronomy, University of British Columbia,
 6224 Agricultural Rd., Vancouver, BC, V6T 1Z1, Canada}

\begin{document}
\date{2006 December}

\pagerange{\pageref{firstpage}--\pageref{lastpage}} \pubyear{2006}

\maketitle

\label{firstpage}
\begin{abstract}
~\citet{leung04} claimed that a previously neglected reheating 
effect makes a small but noticeable change to the process of 
cosmological recombination.  We revisit this effect by considering
a system consisting of both radiation and ionizing gas under adiabatic
expansion.  In the thermal equilibrium limit, due to the huge 
radiation background, only a fraction about $10^{-10}$ of the heat released 
from the recombination of atoms is shared by the matter.  
And in the standard hydrogen recombination calculation, the maximum
fraction of energy lost by the distortion photons through multiple 
Compton scattering is certainly $<10^{-3}$.
Thus this effect is negligible.
\end{abstract}

\section{Introduction}
Detailed calculations of the process through which the early
Universe ceased to be a plasma are increasingly important
because of the growing precision of microwave anisotropy 
experiments.  The standard way to calculate the evolution 
of the  matter temperature during the process of cosmological 
recombination is to consider the expansion of radiation and 
matter separately, and include the relevant interactions,
specifically Compton scattering and photoionization
cooling, as corrections~\citep{peebles71,peebles93,seager00}.  
The matter is treated as a perfect gas 
which is assumed to envolve adiabatically.  
Recently, \citet{leung04} suggested that we need to 
use a generalized adiabatic index, since the gas was initially 
ionized and so the number of species (ion + electron vs atom) 
changes in the recombination process.  In their derivation, they 
considered the effect of photoionization, recombination and 
excitation on the matter, but assumed that the matter was 
undergoing an adiabatic process. However,
an adiabatic approximation for only the ionized matter is not valid 
in this case, because the change of entropy of the matter is not zero.    
Moreover, the photons released from the recombination of atoms mostly 
escape as free radiation, instead of reheating the matter,
since the heat capacity of the radiation is much larger
than that of the matter~\citep[see, for example,][]{peebles93,seager00}.

Here, we try to study this problem in a consistent way by 
considering both the radiation and ionizing hydrogen as components 
in thermal equilibium and under adiabatic expansion.  This is not
{\it exactly} the way things happened during recombination, 
but this will give us the maximum effect of the heat if it 
is all shared by the radiation and matter.

\section{Discussions}
For simplicity, we consider that the matter consists only of 
hydrogen (including helium does not change the physical picture).
By assuming that the radiation field and the matter are in 
thermal equilibrium, the total internal energy per unit mass 
of the system is
{\setlength\arraycolsep{2pt}
\begin{eqnarray}
E &=& \frac{1}{n_{\rm H} m_{\rm H}} \left[ a T^4
+ \frac{3}{2}(n_{\rm H} + n_{\rm e})kT 
+ n_{\rm p} \epsilon_{\rm H,\, ion}
+ \sum_i n_{{\rm H},i} \epsilon_{{\rm H},i} \right] \nonumber \\ 
&=& \frac{1}{m_{\rm H}} \left[ \frac{a T^4}{n_{\rm H}}
+ \frac{3}{2}(1 + x_{\rm e})kT 
+ x_{\rm p} \epsilon_{\rm H,\, ion}
+ \sum_i x_{{\rm H},i} \epsilon_{{\rm H},i} \right],
\label{e1}
\end{eqnarray}} 
\\
where $n_{{\rm H}, i}$ is the number density of neutral atoms in 
the $i$th state, $n_{\rm p}$ is the number density of protons,
$n_{\rm H} \equiv n_{\rm p}+\sum_i n_{{\rm H}, i}$ is the total number 
density of neutral and ionized hydrogens, and $n_{\rm e}$ is the 
number density of electrons.  Additionally $\epsilon_{\rm H,\, ion}$ 
and $\epsilon_{{\rm H},i}$ are the ionization energy for the 
ground state and the $i$th state of hydrogen, respectively,  
the $x$s are the fractional densities 
normalized by $n_{\rm H}$, $T$ is the temperature of the whole system,
$a$ is the radiation constant and $k$ is Boltzmann's constant.

In equation~(\ref{e1}), the first term is the radiation energy, the
second term is the kinetic energy of the matter, and the last two terms
are the excitation energy of the atoms.  Here the energy of the ground
state is set to be equal to zero~\citep{mihalas84}.  No matter 
what energy reference is chosen, the change of energy should 
be the same, i.e.
{\setlength\arraycolsep{2pt}
\begin{eqnarray}
dE &=& \frac{1}{m_{\rm H}} \bigg[ \frac{4 a T^3}{n_{\rm H}}dT 
 - \frac{ a T^4}{n_{\rm H}^2}dn_{\rm H} 
+ \frac{3}{2}(1 + x_{\rm e})k dT  \nonumber \\
&& \quad \quad  + \frac{3}{2}kT dx_{\rm e} 
+\epsilon_{\rm H,\, ion}  dx_{\rm p} 
+ \sum_i  \epsilon_{{\rm H},i} dx_{{\rm H},i} \bigg].
\label{de1}
\end{eqnarray}}
\\
We know that the radiation and matter are not {\it exactly} in thermal 
equilibrium (the two temperatures are not precisely the same) 
during the cosmological recombination of hydrogen, because
the recombination rate is faster than the rate of expansion
and cooling of the Universe. Nevertheless, the radiation
background and matter are tightly coupled and it is a good 
approximation to treat the two as if they were in thermal 
equilibrium~\citep[e.g.][p.232]{peebles71}.
This simple approach allows us to estimate how much of the heat 
released is shared with the radiation field and the matter during the 
recombination of hydrogen.  In the expansion of the Universe, 
the whole system (radiation plus matter) is under an adiabatic process.
However, this is not the case for the ionizing matter on its own, because
the change of entropy of the matter is not zero.  
For an adiabatic process we have 
{\setlength\arraycolsep{2pt}
\begin{eqnarray}
dE &=& P \frac{d\rho}{\rho^2} \\
&=& \frac{1}{m_{\rm H}}\left[ (1+ x_{\rm e})kT + 
\frac{1}{3} \frac{a T^4}{n_{\rm H}} \right] \times \frac{3 dz}{1+z} ,
\label{de2}
\end{eqnarray}}
\\
where $P$ and $\rho$ are the pressure and mass density of the system
and $z$ is redshift.  
By equating equations~(\ref{de1}) and (\ref{de2}), we have
{\setlength\arraycolsep{2pt}
\begin{eqnarray}
\frac{1+z}{T} \frac{dT}{dz} &=&
\frac{3(1+x_{\rm e})kT + \frac{4 a T^4}{n_{\rm H}}}  
{\frac{3}{2}(1+x_{\rm e})kT + \frac{4 a T^4}{n_{\rm H}} }  
 - \frac{1+z}{\frac{3}{2}(1+x_{\rm e})kT + \frac{4 a T^4}{n_{\rm H}}} 
\nonumber \\
&& \times \left[ \frac{3}{2}kT \frac{dx_{\rm e}}{dz}  
 + \epsilon_{\rm H,\, ion}  \frac{dx_{\rm p}}{dz} 
+ \sum_i  \epsilon_{{\rm H},i} \frac{dx_{{\rm H},i}}{dz} 
\right].
\label{f1}
\end{eqnarray}}
\\
In order to see whether we can ignore the radiation field, we need 
to compare the two terms in the denominator, i.e. the radiation 
energy and the kinetic energy of the matter.   If the matter energy
were much greater than the radiation energy, then we would  have  
{\setlength\arraycolsep{2pt}
\begin{eqnarray}
\frac{1+z}{T} \frac{dT}{dz} &\simeq&
2 - \frac{1+z}{\frac{3}{2}(1+x_{\rm e})kT } \nonumber \\
&& \times
\left[ \frac{3}{2}kT \frac{dx_{\rm e}}{dz} 
+ \epsilon_{\rm H,\, ion}  \frac{dx_{\rm p}}{dz} 
+ \sum_i  \epsilon_{{\rm H},i} \frac{dx_{{\rm H},i}}{dz} 
\right],
\label{matter}
\end{eqnarray}}
\\
which is the result given by \citet{leung04}.
However, in the current cosmological model
with $T_0 = 2.725$, $Y_{\rm p} = 0.24$, $h=0.73$ and 
$\Omega_{\rm B}=0.04$~\citep[e.g.][]{spergel06}, we have
{\setlength\arraycolsep{2pt}
\begin{eqnarray}
\frac{E_{\rm matter}}{E_{\rm radiation}} \simeq
\frac{n_{\rm H}kT}{a T^4}
= \frac{n_{\rm H,0}k}{a T_0^3} \simeq 1.6 \times 10^{-10}.
\end{eqnarray}}
\\
So, the radiation energy is {\it much} larger than both the 
kinetic energy of matter and also the total heat released 
during recombination. Hence, we definitely cannot ignore the 
radiation field.  In such a case, the second term in 
equation~(\ref{f1}) is much smaller than the first term, because
{\setlength\arraycolsep{2pt}
\begin{eqnarray}
\frac{(1+z)\epsilon_{\rm H,\, ion}}
{\frac{3}{2}(1+x_{\rm e})kT + \frac{4 a T^4}{n_{\rm H}}} \simeq
\frac{(1+z)n_{\rm H} \epsilon_{\rm H,\, ion}}{a T^4} 
\sim 10^{-6}.
\end{eqnarray}}
\\
Hence the energy change due to the recombination process is taken up mostly 
by the radiation field, since there are many more photons than baryons.
In other words, most of the extra photons (or heat) escape to the 
photon field, with just a very small portion ($\sim$\,$10^{-10}$) 
reheating the matter.  Therefore, the change of the temperature of the 
system can be approximated as
{\setlength\arraycolsep{2pt}
\begin{eqnarray}
\frac{1+z}{T} \frac{dT}{dz} \simeq\ 1 \pm \delta,
\end{eqnarray}}
\\
where $\delta < 10^{-6}$.
This gives us back the usual formula for the radiation temperature,
which is consistent with the result that the matter temperature
closely follows the radiation temperatre~\citep[e.g.][]{peebles68,seager00}.
Leung, Chan \& Chu~(2004) assumed that the extra heat
is shared by the matter only, and hence that the second term of 
equation~(\ref{matter}) is significant, because
{\setlength\arraycolsep{2pt}
\begin{eqnarray}
\frac{(1+z)\epsilon_{\rm H,\, ion}}{(1+x_{\rm e})kT} 
\simeq 6 \times 10^4. 
\end{eqnarray}}
\\
By comparing this and the ratio given in equation~(8), we can see that
the factor is about 10 orders of magnitude larger if we ignore the radiation 
field.  Another way to understand this overestimate is that the 
adiabatic approximation for matter {\it only} is not valid in their 
derivation, because the entropy of the matter is changing 
(i.e. $dS_{\rm matter}/dz > 0$).

The \citet{leung04} paper ignored the last term (the sum 
of the excitation energy terms) in equation~(6), which physically 
 means that there is a photon with energy equal to 
$\epsilon_{\rm H,\, ion}$ ($\sim$13.6eV) emitted when a proton and
electron recombine, and the energy of this distortion photon 
is used up to heat the matter.  This is actually not true for the 
recombination of hydrogen, since there is no direct recombination
to the ground state~\citep{peebles68,zks68,seager00}
and there are about 5 photons per neutral hydrogen atom produced 
for each recombination~\citep{chluba06}.  

Note that what we calculate above is in the thermal equilibrium 
limit and it assumes that all the distortion photons are
{\it thermalized} with the radiation background and the matter.
However, in the standard recombination calculation, 
most of these distortion photons escape to infinity 
with tiny energy loss to the matter through Compton 
scattering.
The maximum fraction of energy loss by the distortion 
photons after multiple scatterings ($\Delta E_{\gamma}/E_{\gamma}$) 
is very low~\citep{switzer05}.  An approximate estimate is
{\setlength\arraycolsep{2pt}
\begin{eqnarray}
\frac{\Delta E_{\gamma}}{E_{\gamma}} 
&\simeq& \frac{\epsilon_{\rm H,\, ion}}{m_{\rm e}c^2} \, \tau \\
&\simeq & \frac{13.6 \, {\rm eV}}{511 \, {\rm keV}} \times 30 
\quad {\rm at} \  z \simeq 1500, \ {\rm when} \ x_{\rm e} \simeq 0.5 
\nonumber \\
& \simeq& 8 \times 10^{-4}, \nonumber
\end{eqnarray}}
\\
where $m_{\rm e}$ is the mass of electron, $c$ is the speed of light and 
$\tau$ is the optical depth.
Therefore, the $\epsilon_{\rm H,\, ion} dx_{\rm p}/dz$ term
is in practise suppressed by 
at least $10^{-4}$ (since $\tau$ decreases when more neutal
hydrogen atoms form at lower redshift).
Hence, although there is {\it some \/}heating of the matter, 
the ratio of the heat shared by the matter and the radiation 
is very small, and the effect claimed by \citet{leung04} is negligible
for the recombination history and also for the microwave
anisotropy power spectra.
\section{Conclusion}
By considering a simple model consisting of the radiation background 
and the ionizing gas under equilibrim adiabatic expansion, 
we show that the effect claimed by \citet{leung04} is hugely overestimated.
The appropriate method for calculating the matter temperature
is to deal with Compton and Thomson scattering between the 
background photons, distortion photons and matter in detail.
In general the Compton cooling time of the baryons off the CMB
is very much shorter than the Hubble time until $z \sim 200$, hence
it is extremely hard for any heating process to make the matter and 
radiation temperatures differ significantly at much earlier times.
\section{Acknowledgements}
We would like to thank Jens Chluba and Henry Ling for very 
useful discussions.
This work was supported by the Natural Sciences and Engineering
Research Council of Canada and a University of British Columbia Graduate
Fellowship.  

\bsp

\label{lastpage}

\end{document}